\documentstyle[aps,twocolumn,epsf]{revtex}

 \def\b {\beta}  \def \d {\delta}
\def\e{\epsilon}  
  \def\la{\lambda}

\def\0{\over } \def\1{\vec } \def\2{{1\over2}} \def\4{{1\over4}}
\def\5{\bar } 
\def\6{\partial }
\def\7#1{{#1}\llap{/}}
\def\8#1{{\textstyle{#1}}} \def\9#1{{\bf{#1}}}
\def\.{\cdot }
\def\^#1{\widehat{#1}}
\def\({\left(} \def\){\right)} \def\<{\langle } \def\>{\rangle }
\def\[{\left[} \def\]{\right]}  

\def\qq{{\bf q}^2}

\def\bq{|{\bf q}|}

\newcommand{\bea}{\begin{eqnarray}}
\newcommand{\beal}[1]{\begin{eqnarray}\label{#1}}
\newcommand{\eea}{\end{eqnarray}}
\newcommand{\be}{\begin{equation}}
\newcommand{\ee}{\end{equation}}
\newcommand{\bel}[1]{\begin{equation}\label{#1}}
\newcommand{\nn}{\nonumber\\}

\begin{document}

\title{Resumming the pressure}

\author{Anton Rebhan}

\address{Institut f\"ur Theoretische Physik, Technische Universit\"at
Wien,\\ Wiedner Hauptstr. 8-10, A-1040 Vienna, Austria}

\date{Talk given at the 5th International
Workshop on Thermal Field Theories\\ and Their Applications, Regensburg,
10--14 August 1998}

\maketitle

\begin{abstract}

The convergence properties of the resummed thermal perturbation series 
for the thermodynamic pressure are investigated by comparison with
the exact results obtained in large-$N$ $\phi^4$ theory and possibilities
for improvements are discussed. By going beyond conventional
resummed perturbation theory, renormalization has to be carried
out nonperturbatively yet consistently. This is exemplified in
large-$N$ $\phi^4_4$ and in a special large-$N$ $\phi^3_6$ model
that mimics QED in the limit of large flavour number.

\end{abstract}


\narrowtext


\section{Introduction}

A few years ago, the authors of Ref.~\cite{QCDP} have accomplished the task
of calculating all contributions to the thermodynamic pressure in QCD
that are accessible to conventional resummed perturbation theory, that is
up to contributions 
from the nonperturbative magnetic sector which involve at least $g^6$.
Dismayingly, the apparent convergence of the perturbation
series up to order $g^5$, rendered in Fig.~1a, proved to be too poor to
warrant its usage at realistic couplings $g\approx2$. In fact,
the convergence is spoiled in particular by the contributions
involving resummation effects, as can be seen in Fig.~1a from
the large jumps whenever a new contribution involving odd powers in $g$
is included. 

Since this problem arises already before one has
to face the nonperturbative nature of the (chromo)\-magnetostatic sector,
which contributes to $O(g^6T^4)$, it may have to do with the constraint
inherent in perturbation theory to truncate everything to 
finite-order polynomials in the coupling (up to possible logarithms).
This is necessary in view of the needs of the renormalization process,
but it certainly discards a large portion of the resummation effects
which are in principle contained in low-order diagrams.

The replacement of truncated perturbation series by Pad\'e approximants,
i.e.\ by perturbatively equivalent rational functions, as proposed
in Ref.~\cite{KP} can be considered
as a very simple guess what the discarded terms may have been.
At any rate, it gives a rough idea as to their importance. In Fig.~1b
the result for the pressure is transformed by appropriate Pad\'e approximations,
and it does have a big effect for larger coupling. The apparent convergence
is somewhat improved, but at $g\approx2$ the result is still inconclusive;
actually, the final result including the 5th order contributions 
looks even worse than before as it tends to values larger 
than the free-pressure one.


\begin{figure}
\vspace{-4mm}
\epsfxsize=7cm
\centerline{\epsfbox{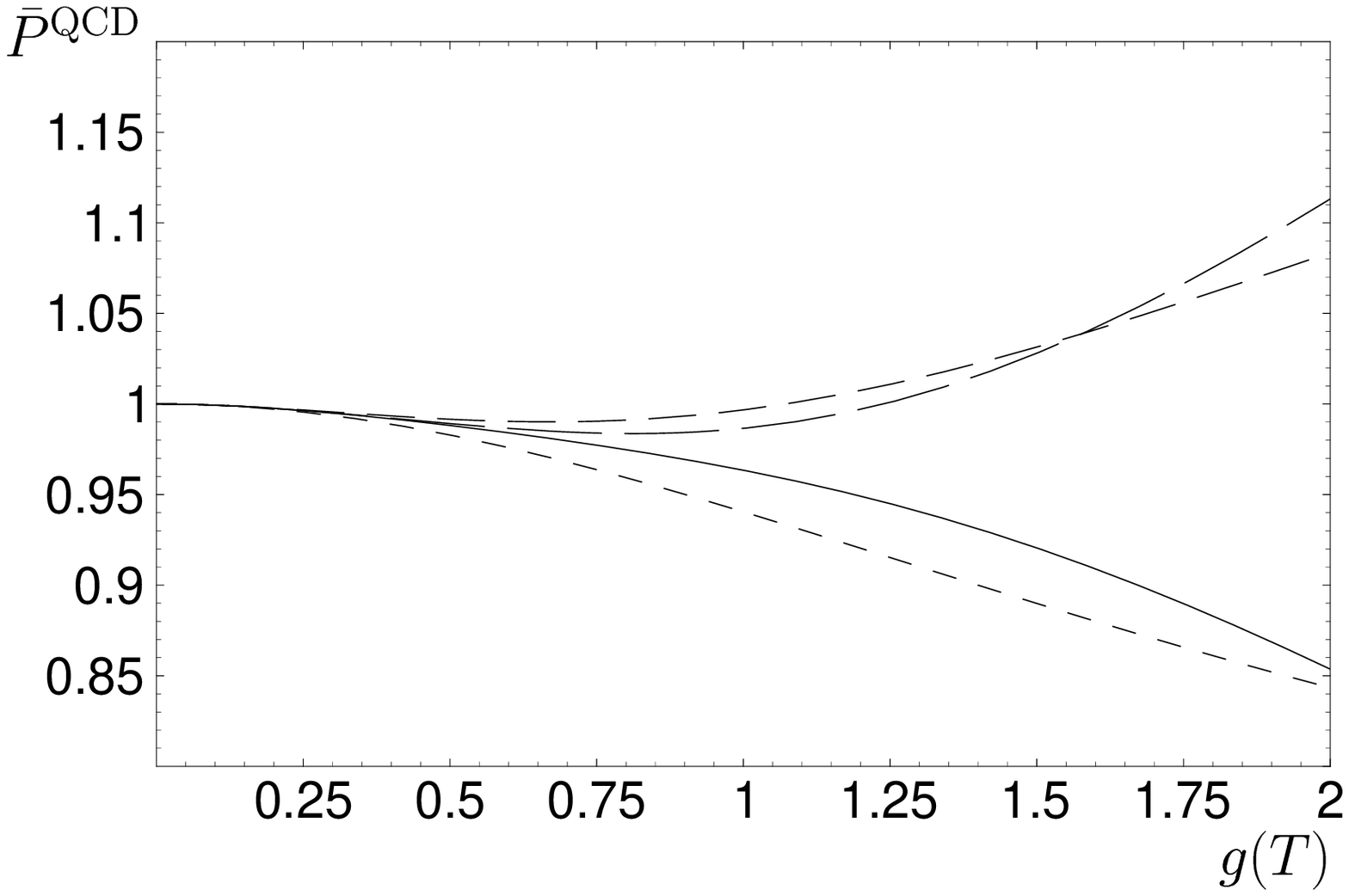}}
\centerline{(a)}\epsfxsize=7cm
\centerline{\epsfbox{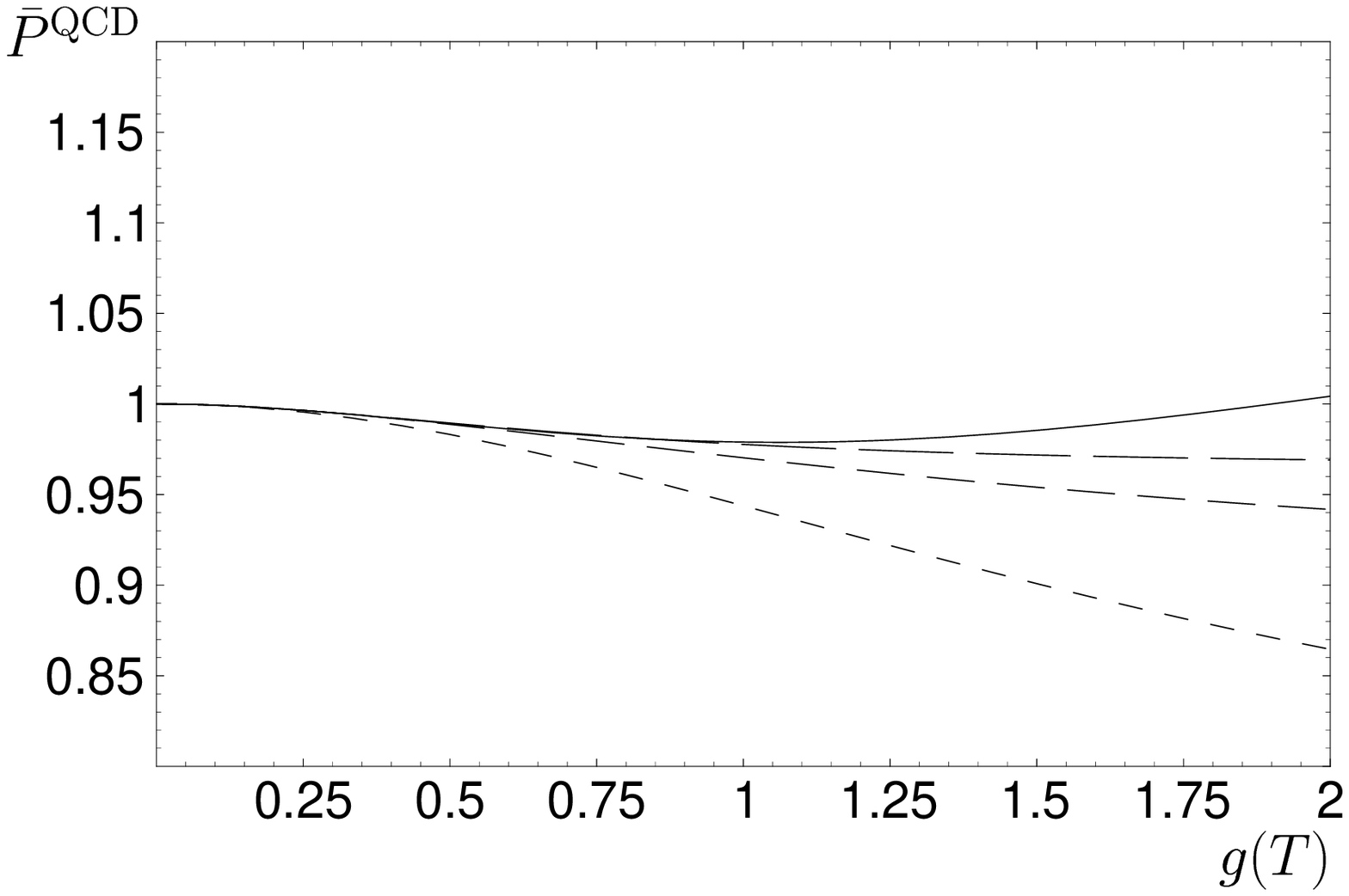}}
\centerline{(b)}
\vskip4mm
\caption{a) The perturbative results for the pressure of QCD($N_f=3$)
up to order $g^5$ (full line). Short, medium, and long dashes give the results
up to $g^2$, $g^3$, and $g^4$, respectively. b) The corresponding
Pad\'e approximants $[0,2]$, $[1,2]$, $[2,2]$, and $[3,2]$.}
\label{fig1}
\end{figure}

In this talk, I shall discuss the convergence properties of resummed thermal
perturbation theory in simple solvable but nontrivial
cases. The starting point will be an exact nonperturbative formula for the
full pressure which itself has the form of a one-loop integral.
Its evaluation and renormalization will be carried out first in
the large-$N$ limit of scalar $\phi^4$-theory, which is simplified
by the fact that the self-energy is momentum-independent, and
then in a special large-$N$ $\phi^3_6$ model
that mimics QED in the limit of large flavour number, which brings
in some of the complications coming from momentum-dependent self-energies.
Moreover, these toy models have some noteworthy features which
also make them interesting on their own.

\section{A one-loop formula for the nonperturbative pressure}

In Ref.~\cite{DHLR1} we found that at the cost of introducing a
further integration with respect to the bare masses of a field
theoretical model the full thermodynamic pressure can be expressed
as a one-loop integral involving the full self-energy. This follows
from the observation that
\be
{\6\0\6m_0^2}P(T)={1\0i\b V}\left\< \2 \int_C d^4 x \phi^2(x) \right\>
=-\2 \< \phi^2(0) \>
\ee
where $C$ is the Keldysh contour of real-time finite-temperature
field theory \cite{LeB}\footnote{For a concise review see the
lectures of Peter Landshoff at this meeting\cite{PL}},
and from the requirement that the pressure
$P(T)-P(0) \to 0$ when all masses are sent to infinity.\footnote{This
approach has obvious similarities with the so-called
exact renormalization-group approach of Ref.~\cite{W}} 
Thus
\beal{Pr}
&&P(T)-P(0) = \2 \int_{m_0^2}^\infty dm_0'^2 \< \phi^2(0) \>_{T,m_0'}-
\biggl( T=0 \biggr) \nn
&&=\2 \int_{m_0^2}^\infty dm_0'^2 \int {d^nq\0(2\pi)^n}
\[ D^{12}_T(q;m'_0)-D^{12}_0(q;m'_0) \] \nn
&&=-\2 \int_{m_0^2}^\infty dm_0'^2 {\rm Im} \int {d^nq\0(2\pi)^n}
\biggl\{ {1+2n(q^0)\0q^2-m_0'^2-\Pi_T(q^0,\qq;m_0')}\nn
&&\qquad\qquad\qquad\qquad\qquad\qquad-{1\0q^2-m_0'^2-\Pi(q^2;m_0')}
\biggr\}
\eea
where $D^{ab}$ denotes the real-time matrix propagator and $\Pi$
is obtained by diagonalization of the self-energy matrix.
This can be easily generalized to fermions and in principle even to
(gauge-fixed) gauge theories\cite{DHLR1}. While it is true that
the introduction of auxiliary masses generally
spoils both gauge and BRS symmetries, the combinatorics of the
loop expansion which is reshuffled by the mass integration in
Eq.~(\ref{Pr}) does not depend on these symmetries. So although
their loss would be a high price in intermediate steps, the full
integral depends only on the physical masses (which are
zero for gauge bosons).

One noteworthy aspect following from the essential one-loop nature
of the above (exact) formula is that it appears to be manifestly infrared-finite
in all dimensions $n>3$, irrespective of whether or not $\Pi_T$
gives rise to screening (or even divergences). So this
seems to be a good new starting point for some reorganization of
thermal perturbation theory---which would have to see that $\Pi_T$
when obtained in some approximation
is not expanded out again from the denominator in Eq.~(\ref{Pr}). Clearly, this would require a correspondingly
reorganized renormalization procedure. So far, everything has been
written down in terms of unrenormalized quantities only.

\section{Nonperturbative renormalization in the example of
large-$N$ $\phi^4_4$ theory}

The large-$N$ limit of a scalar O($N$) model with interaction
Lagrangian
\be
{\cal L}_I=-{\lambda _0\over 4!}{3\0N+2} 
\left (({{\vec\phi }}(x))^2\right )^2
\ee
can be solved exactly \cite{DJ,BM}. It can also be viewed as a
certain approximation to ordinary ${\lambda _0\over 4!}\phi^4$-theory,
in which only those diagrams are kept which have a topology that
corresponds to the leading term of a $1/N$-expansion. This infinite
set of diagrams is alternatively called Hartree-Fock, super-daisy, cactus,
or foam \cite{DHLR2} diagrams.

In this approximation, the Schwinger-Dyson equation for the self-energy
does not involve vertex functions and is given by a one-loop equation.
Its renormalization at $T=0$ gives
\be
m^2=m_0^2+\la _0M(m^2)
\ee
with
\be
M(m^2)=\2\int {d^nq\over(2\pi)^n}{i\over q^2-m^2+i\e}
\ee

Similarly, coupling constant renormalization is given by
\be
\la=\la _0+\la _0\la   M'(m^2)
\ee
when renormalizing at zero momentum.
In four dimensions, this leads to the problem of triviality if one
requires both $\la_0$ and $\la$ to be positive, for then $\la\to0$
as $n\to4$. Keeping $\la>0$ is only possible by using a cut-off or
by accepting $\la_0<0$. In the latter case one finds that the scattering
amplitude has a tachyonic pole, but this occurs at $s=s_{\rm tach}\approx
-m^2 e^{32\pi^2/\la}\approx -m^2 10^{137/\la}$, which is exponentially
huge for reasonably small coupling. Either way we can accept
this theory as an effective theory for momenta and temperatures $\ll \sqrt
s_{\rm tach}$.

At finite temperature, the Schwinger-Dyson equation leads to
a ``gap'' equation of the form
\be
m^2+\delta m^2=m_0^2+\la _0 M_T(m^2+\delta m^2)\ee
with
\bea
&&M_T(m^2)-M(m^2)\equiv N_T(m^2)\nn&&=\int {d^nq\over(2\pi)^n}2\pi\delta ^+(q^2-m^2){1\over e^{q^0/T}-1}\eea

Elimination of the unrenormalized parameters in favour of the
renormalized ones yields
\be
\delta m^2(m^2,T)=\la [\hat M(m^2,\delta m^2)
+N_T(m^2+\delta m^2)]
\ee
with
\bea
&&\hat M(m^2,\delta m^2)\nn
&&\quad=M(m^2+\delta m^2)-M(m^2)-\delta m^2M'(m^2)
\eea

The function $\hat M$ is formally of order $\la^2$, i.e.~of the
same order as 3-loop contributions to the pressure, and it has
been occasionally missed in the literature \cite{DJ}. It exhibits
a nontrivial interplay of the thermal mass correction $\delta m^2$
with the zero-temperature UV-divergent quantities $M(m^2)$ and $M'(m^2)$
appearing in the 
counter-terms.

Rewriting also the pressure in terms of renormalized quantities, we
arrive at the remarkable formula
\bel{Pri}
P(T)-P(0)=\int_{m^2}^\infty dm'^2{\delta m^2(m'^2,T)
\over\lambda(m'^2)} 
\ee
where $\lambda(m'^2)$ is a running coupling that is equal to $\la$ when
$m'^2=m^2$; $\la_0$ is kept fixed throughout.

Fortunately, Eq.~(\ref{Pri}) can be integrated, yielding\footnote{An
explicit expression of the pressure in large-$N$ $\phi^4$ has been
obtained previously in Ref.~\cite{ACP}, though 
their formula does not satisfy the physically-important constraint
that the pressure vanishes when the mass is infinite.}
\beal{Pris}
P(T)-P(0)&=&P_T^{\rm free}(m^2+\delta m^2)+ \2 {(\d m^2)^2\0 \la}\nn&&-
\sum_{n=3}^\infty {1\0n!}M^{(n-1)}( m^2)(\d m^2)^n
\eea
While the second term on the right-hand-side can be identified
as an interaction contribution, the subsequent terms again come from
the thermal mass shift in zero-temperature integrals.

While being UV-finite, the above equation becomes IR-divergent when
$m\to0$. This is not a problem specific to finite temperature, but
comes from the breakdown of the on-shell renormalization scheme that
we have used so far. In the limit $m\to0$ one can switch to
an off-shell ($\overline{\hbox{MS}}$) scheme through
\be
\la^{-1}=\bar\la^{-1}+{1\032\pi^2}\log{\bar\mu^2\0m^2}
\ee
which simplifies
\be
\hat M \to \bar M(\bar\mu^2,\delta m^2)=
{1\032\pi^2}\delta m^2\;(\log{\delta m^2\0\bar\mu^2}-1)
\ee

It is now only a matter of simple numerical integrations to
calculate $\d m^2$ and the resulting pressure.

Somewhat surprisingly, there is not one but two solutions for
$\d m^2$ for small values of the coupling, and none for $\bar\la
> \bar\la_{\rm crit}$, whose value depends (in a renormalization-group
invariant manner) on the renormalization scale $\bar\mu$.
However, the higher values of $\d m^2$ that correspond to one and the
same $\bar\la$ are close to the tachyonic scale $\surd{s_{\rm tach}}$
(see Fig.~2),
which we have agreed to ignore. In order that $\surd{s_{\rm tach}}$
be exponentially far away, we need $\bar\la(T) \ll 10^2$, which
also cuts out the case of no solution for the thermal mass and
therefore for the pressure.\footnote{This is in accordance with the
findings of Ref.~\cite{BM}}

\begin{figure}\epsfxsize=7cm
\centerline{\epsfbox{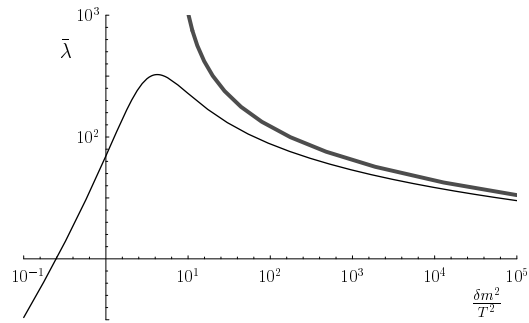}}
\vskip4mm
\caption{The relation between solutions for $\d m^2$ and the
renormalized coupling $\bar\la$; the heavy line marks where $\delta m$ would
become equal to the (modulus of the) tachyon mass.}
\label{fig2}
\end{figure}

\section{Comparison of a nonperturbative result with 
perturbation series}

Besides its pedagogical value, the above 
nonperturbative result can be put to use
to study the properties of (truncated) series expansions in
$\bar\la(\bar\mu)$. The first few terms in the result for
the pressure read\footnote{Up to and including order $\bar\la^{3/2}$,
where there is no difference between the foam-diagram subset and
the full set of diagrams, this agrees with the full
$N=1$ result obtained using resummed perturbation theory in Ref.~\cite{PS}.
}
\bea
&&{P(T)-P(0)\0T^4}={\pi^2\090}-{\bar\la\01152}+{\bar\la^{3/2}\0 576\pi\sqrt6}
\nn&&\qquad\qquad
-(6-\gamma-\log{\bar\mu\04\pi T}){\bar\la^2\018432\pi^2}
\nn&&\qquad
+(3-2\gamma-2\log{\bar\mu\04\pi T}){\bar\la^{5/2}\012288\pi^3\sqrt6}
\nn&&\quad
-\( (6-\gamma-\log{\bar\mu\04\pi T})^2-30+{\zeta(3)\036} \)
{\bar\la^3\0294912\pi^4}
\nn&&\qquad\qquad
+O(\bar\la^{7/2})
\eea

Comparing truncated series expansions like the above with the full
nonperturbative result, we can investigate the convergence properties
of an expansion in powers (and logarithms) of $\bar\la^{1/2}$.
It turns out that these depend strongly on the ratio of the
arbitrary renormalization scale $\bar\mu$
to temperature.

In Fig.~3 we juxtapose the exact and the perturbative
result for the pressure 
\hbox{$[P(T)-P(0)]$}
to its ideal-gas value $\pi ^2T^4/90$,
including successively up to 10 terms beyond the
leading one. We choose various 
values of the
renormalization scale $\bar\mu$,  but 
for ease of comparison
in each case we plot against $\bar\la$
evaluated for $\bar\mu=T$.

This shows that when $\bar\mu$ is very different from $T$, the
convergence of the series deteriorates significantly. For $\bar\mu=100T$
(Fig.~3a), the truncated series develop oscillatory behaviour for
larger values of the coupling, whereas for $\bar\mu={1\0100}T$,
the perturbative results fail to improve with increasing order
at roughly the same value of the coupling, albeit in a more
peaceful manner. With $\bar\mu\approx 2\pi T$, the perturbation series
tolerates the largest coupling strength, but it is evident that
the perturbative results cannot describe more than a few percent
deviation from the free pressure value.

\begin{figure}\epsfxsize=7cm
\centerline{\epsfbox{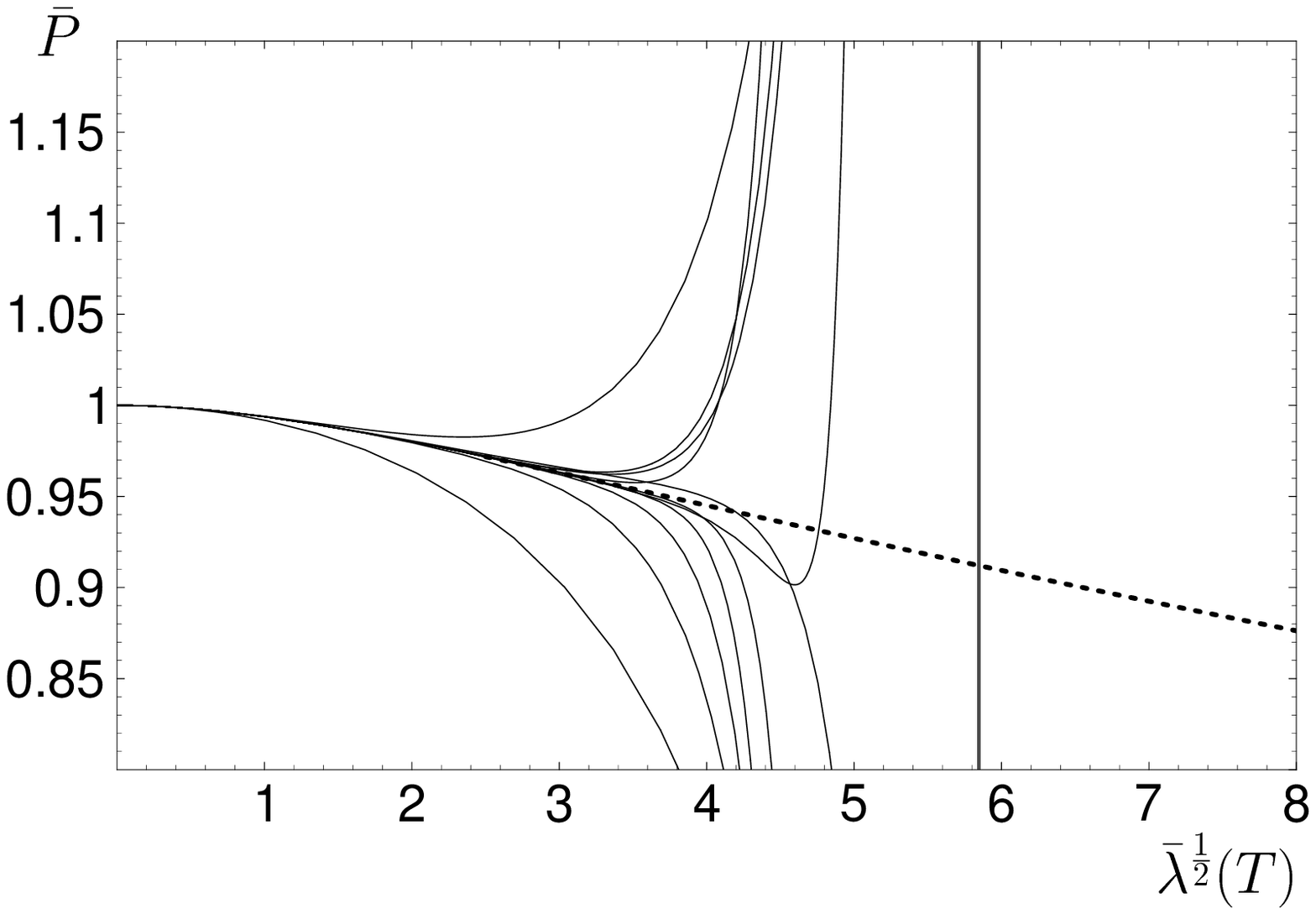}}
\centerline{(a)}
\epsfxsize=7cm
\centerline{\epsfbox{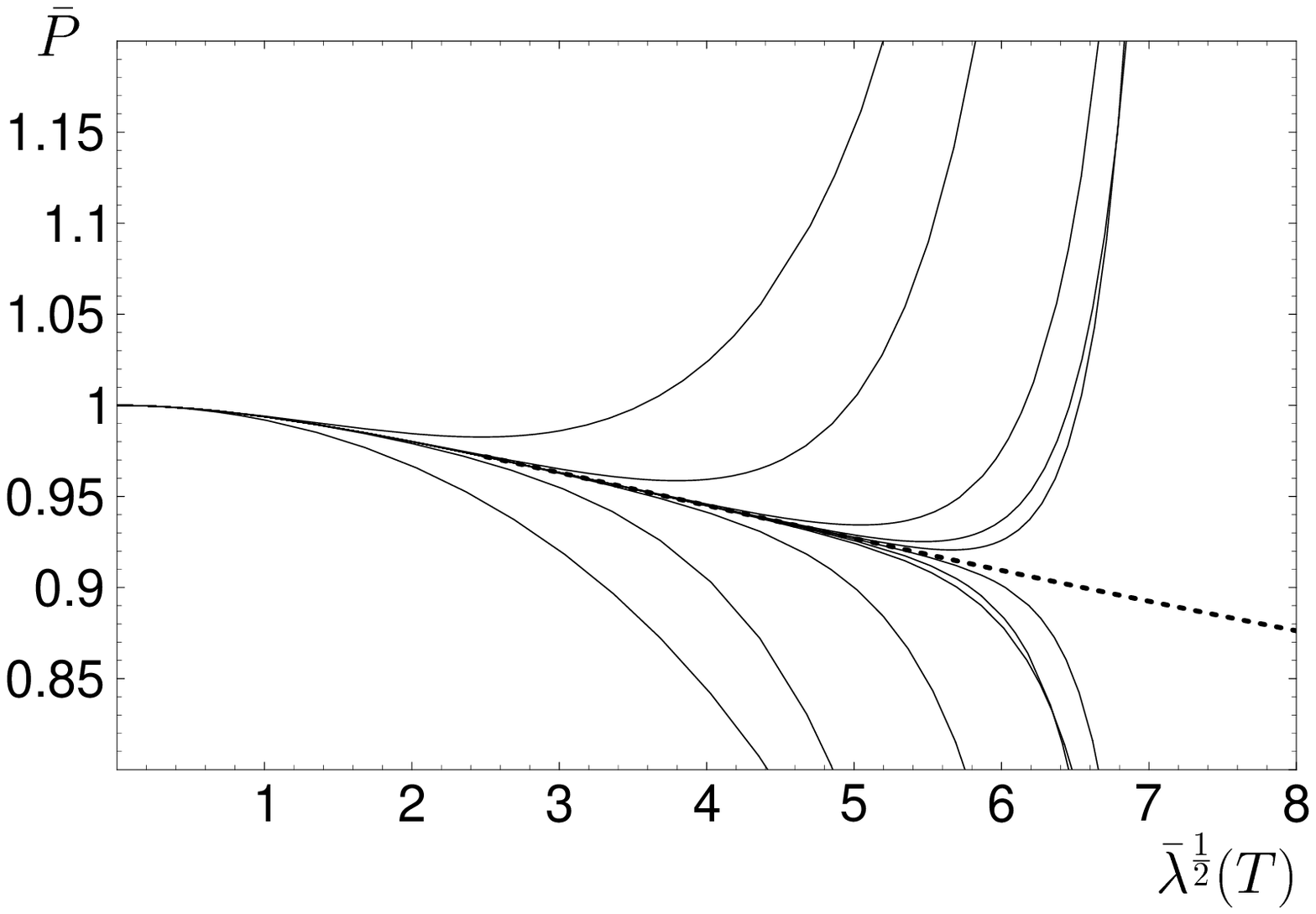}}
\centerline{(b)}
\epsfxsize=7cm
\centerline{\epsfbox{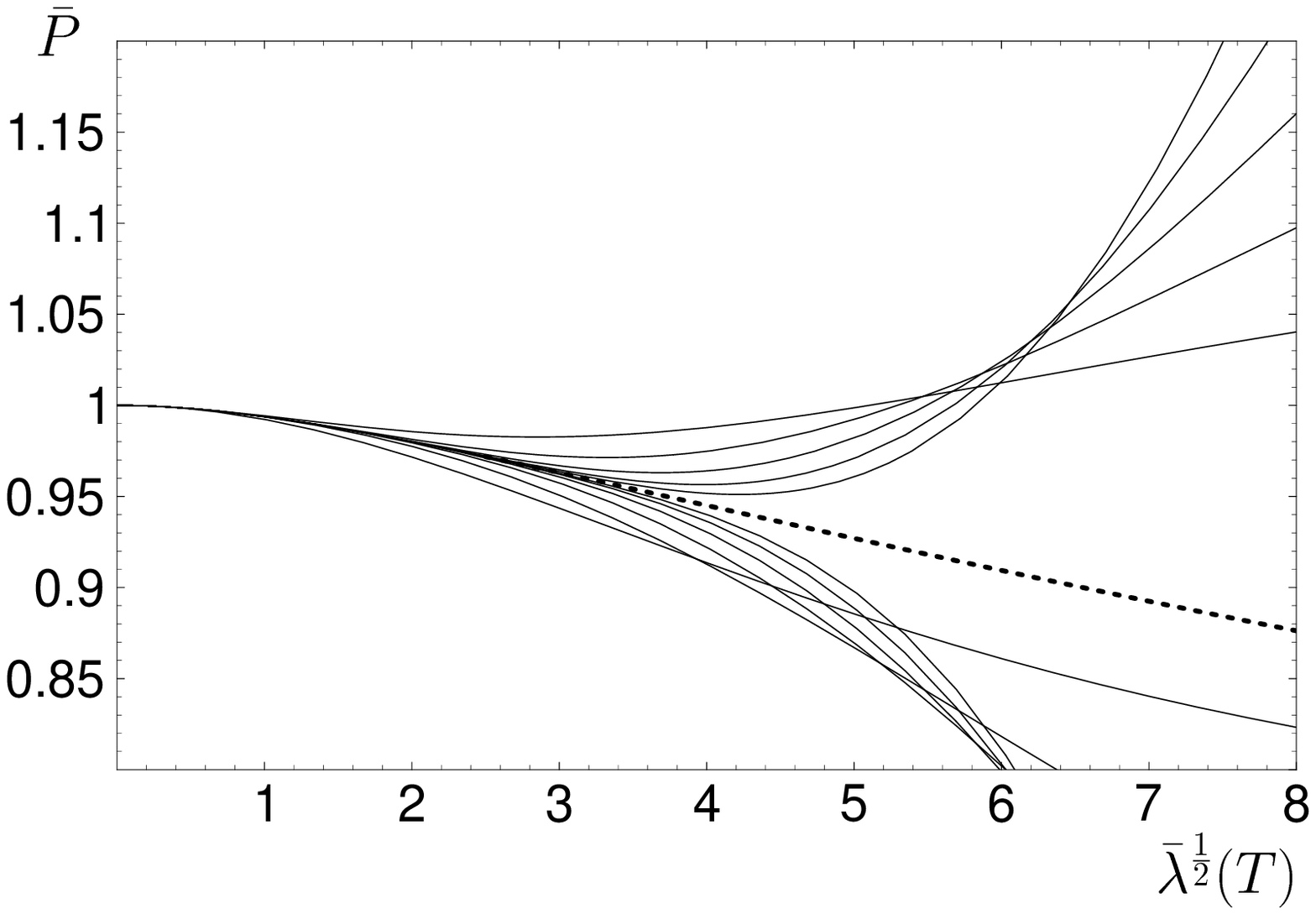}}
\vskip6mm
\caption{A comparison of the perturbative results for $\bar P\equiv
[P(T)-P(0)]/{\pi^2T^4\090}$
as a function of ${\bar\la^{1/2}(T)}$
up to 12th order for different choices of the
renormalization scale: a) $\bar\mu=100T$, b) $\bar\mu=2\pi T$, 
c) $\bar\mu={1\0100}T$.}
\label{fig3}
\end{figure}

In Fig.~4 the results of a Pad\'e improvement is displayed. This is
seen to work surprisingly well except for those cases where the
Pad\'e approximant develops a pole beyond which the result appears to
be off by a constant.

In QCD, the Pad\'e approximants did not work nearly as well
(cf.\ Fig.~1b). This might
have to do with the absence of $\log(\bar\la)$-terms in the
large-$N$ limit of our model. In QCD similar logarithmic terms are present
and they are treated like constants when building the Pad\'e
approximants.

\begin{figure}\epsfxsize=7cm
\centerline{\epsfbox{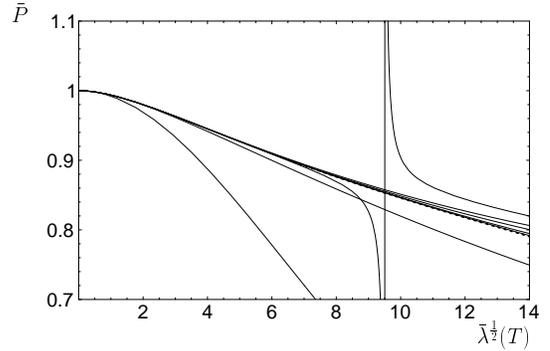}}
\vskip6mm
\caption{As in fig.~3b, but with the perturba\-tive power series replaced
by Pad\'e approximants $[0,2]$, $[1,2]$, $[2,2]$, $[2,3],\ldots,[4,4]$. 
They quickly
converge to the exact result with the exception of $[3,3]$, which has
a pole.
}
\label{fig4}
\end{figure}

Let us finally try out a completely different expansion. If we
a posteriori expand everything in powers of $\d m/T$ rather than
$\bar\la$, we find that the convergence properties are excellent
so that substantial deviations from the free pressure results can
be covered with only a few terms kept (Fig.~5). The reason for this
is that the high-temperature expansion has a convergence radius of
$\d m/T=2 \pi$, a value which, at least in our model, is always
far away for all $\bar\la \ll \bar\la_{\rm crit}$.

\begin{figure}\epsfxsize=7cm
\centerline{\epsfbox{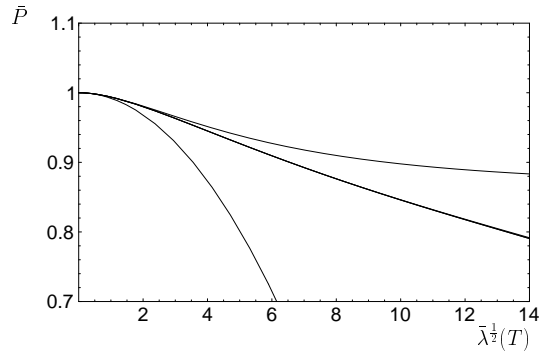}}
\vskip6mm
\caption{The pressure with solutions of the thermal mass equation truncated at leading order in $\delta m/T$ (top curve) and 
next-to-leading order (bottom curve). 
All higher approximations are virtually
indistinguishable from the exact solution,
and they collectively form the middle line, which broadens
only at the highest values of $\bar\la$.}
\label{fig5}
\end{figure}

\goodbreak
This, perhaps merely tantalizingly, indicates that a reorganization of
thermal perturbation theory as a series in $\d m /T$ rather than $g$
should lead to dramatic improvements. In this connection
see also the proposal of a ``screened perturbation theory'' in
Ref.~\cite{KPP} and the contribution of S.\ Leupold at this conference
\cite{SL}.

\section{Nonperturbative renormalization in the example of
large-$N$ $\phi^3_6$ theory}

Large-$N$ $\phi^4$-theory was exceptionally simple in that the
self-energy is momentum independent. In order to learn how to
deal with the complications from a momentum-dependent self-energy
and the need of wave function renormalization we now consider
a scalar $\phi^3_6$ theory, in which a {\em scalar} isovector $A$
is coupled to $N\to\infty$ {\em scalar} isodoublets $\psi$ according to
\cite{BLNR}
\be
{\cal L}_I=
-{g\0\sqrt{N}}A^a\psi _{r} ^{\dag}\tau ^a\psi _{r},\qquad r=1,\ldots,N
\ee
so that this model mimics large-$N_f$ QED, but without the
complications of spinor and vector boson fields.
The masses of the
fields $A$ and $\psi$ are denoted by $m$ and $M$, respectively,
and we require $m<2M$ in order to have stable photons.

The same trick as in Sect.~2 can be used to derive a one-loop formula
for the full pressure, but it suffices to differentiate and integrate
with respect to the ``photon'' mass $m$, for $m\to\infty$ switches off
all interactions. Thus
\beal{Pr3}
&&P(T)=P^{\rm free}_{(\psi)}(T,M_0^2)-{3\02}{\rm Im}\int_{m_0^2}^\infty dm_0'^2
\int{d^{6-2\e}q\0(2\pi)^{6-2\e}}\biggl\{ \nn
&&\qquad{1+2n(q^0)\0q^2-m_0'^2-g_0^2\pi_T(q^0,\qq)}-{1\0q^2-m_0'^2-g_0^2
\pi(q^2)}
\biggr\}
\eea
where $\pi_T(q^0,\qq)$ and $\pi(q^2)$ are the finite-temperature and
zero-temperature photon self-energies (with two powers of the coupling
factored out), respectively.

In the limit of $N\to\infty$, the interaction pressure is of order $N^0$,
and to this order there are no corrections to the internal ``electron''
lines in the one-loop photon self-energy, because the electron
self-energy is of order $N^{-1}$. So the latter can be ignored except
in $P^{\rm free}_{(\psi)}(T,M_0^2) \sim N$, where the renormalization of
$M_0$ contributes to the interaction pressure, and is in fact
essential to make the latter finite.

The UV divergences in $\pi(q^2)$ have to be eliminated by mass and
wave-function renormalization. Choosing on-shell renormalization we
have
\bea
m'^2&=&m_0'^2 + g_0^2 \pi(m'^2)\\
g^2(m'^2)&=&Z_2(m'^2)g_0^2\\
Z_2(m'^2)&=&{1\01-g_0^2\pi'(m'^2)}={\6m'^2\0\6m_0'^2}\\
\bar\pi(q^2)&=&\pi(q^2)-\pi(m'^2)-(q^2-m'^2)\pi'(m'^2)\\
\bar\pi_T(q^0,\qq)&=&\pi_T(q^0,\qq)-\pi(m'^2)-(q^2-m'^2)\pi'(m'^2)
\eea

In Eq.~(\ref{Pr3}) the mass integration can be carried out, which is
trivial before renormalization and only slightly less so after,
with the result
\beal{Pri3}
&&P(T)^{\rm int}\equiv P(T)-P(T)\Big|_{g=0}\nn
&&\quad=-{3\02}{\rm Im}\int{d^6q\0(2\pi)^6}\biggl[
2n(q^0)\log{g^2\bar\pi_T(q^0,\qq)+m^2-q^2\0m^2-q^2}\nn
&&\quad\qquad\qquad\qquad\qquad+\log{g^2\bar\pi_T(q^0,\qq)+m^2-q^2\0g^2\bar\pi(q^2)+m^2-q^2}\nn
&&\quad\qquad\qquad\qquad\qquad-{g^2\hat\pi_T(q^0,\qq)\0g^2\bar\pi(q^2)+m^2-q^2} \biggr]
\eea
The last term in the square brackets comes from
$P^{\rm free}_{(\psi)}(T,M_0^2)-P^{\rm free}_{(\psi)}(T,M^2)$
and is necessary to remove the UV divergence contained in the second
term. There is a subtlety involved here in that
$\hat\pi\equiv \bar\pi_T-\bar\pi$ except for spacelike momenta; for those
this equality holds for the real parts only, while the
imaginary parts differ by contributions that vanish exponentially
in the ultraviolet. This again provides an example of
a nontrivial effect of zero-temperature renormalization
in thermal quantities.

Beyond the large-$N$ limit, the renormalization of Eq.~(\ref{Pr3})
becomes much more involved and requires a careful treatment of
overlapping divergences, see Ref.~\cite{BLNR}.

In the large-$N$ limit, we had to consider only
one-loop (but this time non-local!) contributions to 
the self-energies---still,
Eq.~(\ref{Pri3}) is clearly a nonperturbative result
involving arbitrarily high powers of $g$.

It is in fact a result which could not have been obtained by
the methods of hard-thermal-loop resummation. The hard-thermal-loop
approximation to $\pi_T$ turns out to be identical in form
to the longitudinal component of $\Pi_{\mu\nu}^{\rm HTL}$ in 
four-dimensional gauge theories, but with a reversed over-all sign:
\bea
&&\bar\pi _T^{\rm HTL} (q^0,{\bf q}^2)=\nn
&&\qquad\qquad-{T^2\024\pi} \(1-{q_0^2\0\qq}\)
\[ 1- {q_0\02\bq} \log {q_0+\bq\0q_0-\bq}\]
\eea
At $q^0$ this leads to a screening mass squared with a wrong sign,
giving rise to a spacelike pole in the hard-thermal-loop resummed propagator.
This has been noted before in $\phi^3_6$-theories \cite{Pis3}, and it
is similar to what occurs in the hard-thermal-loop graviton
propagator when evaluated (inconsistently) on flat space\cite{Grav},
where this is interpreted as the Jeans instability of a gravitating medium.

In our scalar case, it is a reflection of the unboundedness of the potential
from below. For small temperatures, the mass term for the photon
stabilizes the system, but thermal mass corrections lead to a
diminished total mass up to a point $T=T_{\rm crit}$ where
the screening mass vanishes. 
Beyond this point, an instability develops, because thermal fluctuations
become able to surmount the barrier provided by the zero-temperature
mass term in the potential, and this is reflected by the
appearance of a ``tachyonic'' screening mass (Fig. 6). 

Correspondingly, the pressure
(\ref{Pri3}) is well-defined and real only for $T\le T_{\rm crit}$.
A hard-thermal-loop approximation fails because it approaches the
problem from the wrong ``side''.
\begin{figure}
\centerline{\epsfxsize=5.5cm\epsfbox{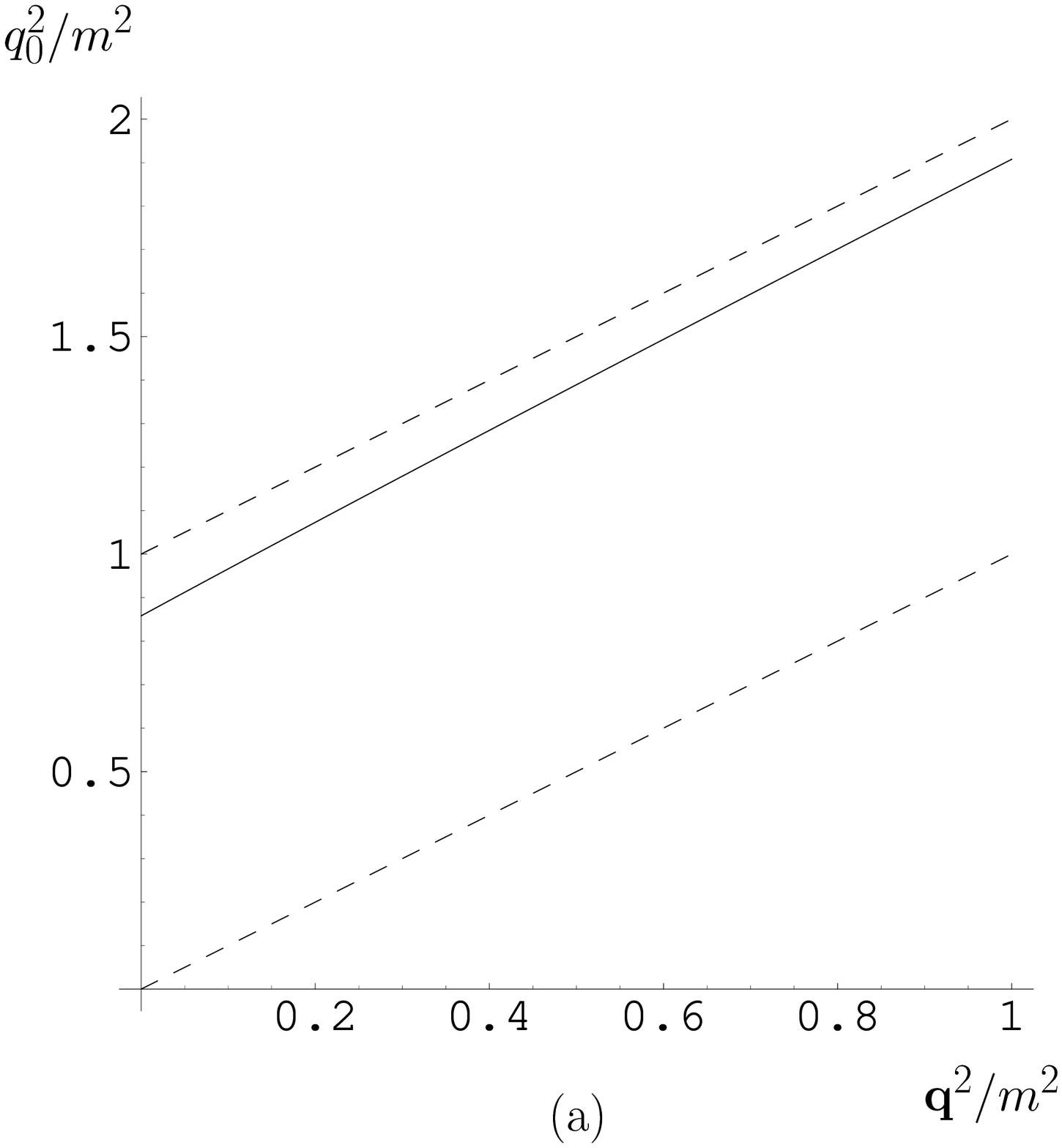}}
\centerline{\epsfxsize=5.5cm\epsfbox{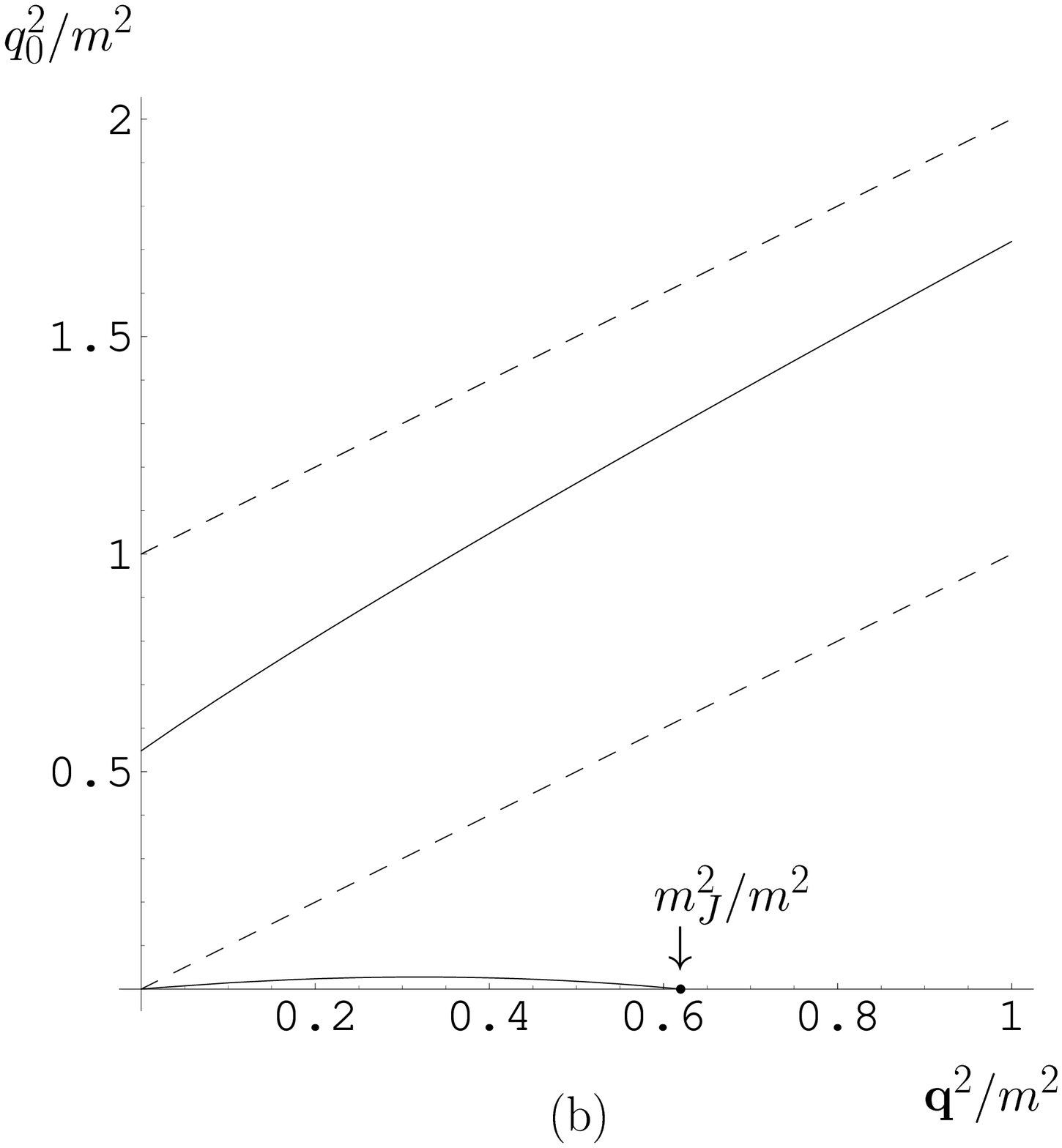}}
\vskip2mm
\caption{The location of poles in the ``photon'' propagator
for $g=10$ at $T=m=m_1=m_2$ (a), which is below  the critical temperature, and 
at $T=1.5m$ (b), which is above it. The two
dashed lines mark the light cone and the zero-temperature mass
hyperboloid $q_0^2=\qq+m_2^2$. The  line between them
is the location of the poles in the thermal propagator. In (b)
the second full line below the light cone marks zeros of the
real part of the inverse propagator, which correspond to
poles of the propagator only at its intersections with $q_0=0$
because of large Landau damping for $q_0>0$ and $q^2<0$.}
\label{fig6}
\end{figure}

On the other hand, the full nonperturbative $N\to\infty$ result (\ref{Pri3}) 
can be evaluated numerically (if tediously) by a couple of nested
numerical integrations ($\pi_T$ cannot be expressed in
terms of elementary functions, but has a rather involved
analytic structure).

In Fig.~7, the results of such a numerical evaluation is given for
three values of the coupling $g$ and $m=M$, and it is compared with the
strictly perturbative $g^2$-part of the potential (without
hard-thermal-loop resummation).

The nonperturbative pressure exists up to the point $T=T_{\rm crit}
\approx \sqrt{24\pi}m/g$. Presumably there is a finite but
complex analytic
continuation beyond this point, but our derivation is in terms of
a manifestly real quantity, which becomes infrared singular
for $T> T_{\rm crit}$. Right at the critical temperature, however,
where (\ref{Pri3}) is still well-defined,
we are as close as possible to the behaviour of four-dimensional gauge
theories in that there is a vanishing screening mass, as is the
case for magnetostatic modes in QED.

\begin{figure}\epsfxsize=9cm
\centerline{\epsfbox{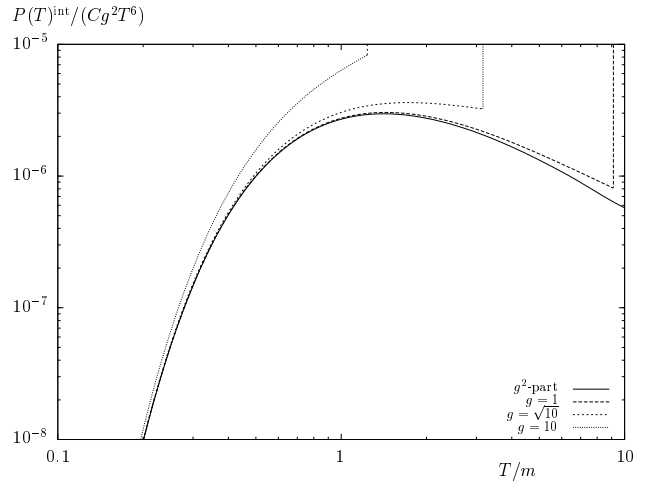}}
\vskip1mm
\caption{Numerical evaluation of the interaction
pressure (\ref{Pri3}) for three
values of $g$
in comparison with the leading perturbative contribution.}
\label{fig7}
\end{figure}

\section{Conclusion}

At least as concerns the thermodynamic pressure of hot QCD,
conventional (hard-thermal-loop) resummed perturbation theory,
even after
including everything that can be
treated perturbatively, falls fails to give reliable results
for moderately large coupling.
The perturbative series
shows a surprisingly poor apparent convergence, far worse than
usually the case for the range of coupling considered, with the
consequence that only a few percent of deviation from the ideal-gas
result can be described.

It could well be that this bad behaviour comes from a certain incompleteness
of the resummation of hard thermal loops, because finally everything
has to be broken down to a truncated series, that is, a polynomial
in $g$ and $\log(g)$.

Pad\'e approximants are the simplest possibility to replace
the latter by functions that are more likely to fit the actual
behaviour at larger values of $g$, but in QCD they only lead
to a marginal improvement, if any.

In order to help prepare the (uncertain) way to alternative resummation
schemes, we have considered simple scalar theories. Going
beyond a strictly perturbative scheme in the coupling involves
a correspondingly nonperturbative renormalization. In the simple
models that we have considered we have demonstrated the importance
of a consistent renormalization scheme. 

In large-$N$ $\phi^4$ theories we have been able to compare
thoroughly the convergence behaviour of an expansion in the
coupling with the full nonperturbative result. We have seen
the importance of the choice of renormalization scale, but
have found that the convergence of these expansions is
limited to small coupling and that the range of the latter
increases rather slowly with the order of perturbation theory.
In contrast to QCD, Pad\'e approximants worked surprisingly well,
but this may be specific to the model considered, for which 
the pressure fortuitously
did not involve logarithms in the coupling constant.
An exceedingly good approximation was obtained by an expansion
in terms of $\d m/T$ in place of $g$, but it is of course totally
unclear how to implement such a scheme in more complicated theories.

\acknowledgments

The work presented here was done in collaboration with P. Landshoff
(to whom I am especially grateful for a critical reading of this
write-up), D. B\"odeker, I. Drummond, R. Horgan, and
O. Nachtmann. It has been partially supported by the Austrian 
``Fonds zur F\"orderung der wissenschaftlichen
Forschung (FWF)'', project no. P10063-PHY, and
by the Jubil\"aumsfonds der
\"Osterreichischen Nationalbank (project no 5986).
I would also like to thank U. Heinz and the local organizers of
the ``5th International Workshop on Thermal Field Theories and
Their Applications'' for all their efforts.


\end{document}